\documentclass{aastex}
\usepackage{spr-astr-addons}
\usepackage{longtable}
\begin{document}
\title{Dependence of Fanaroff--Riley break of radio galaxies on luminosity and redshift}
\shorttitle{Dependence of Fanaroff--Riley break on luminosity and redshift}
\shortauthors{Singal and Rajpurohit}

\author{Ashok K. Singal\altaffilmark{1}} \and \author{Kamlesh Rajpurohit\altaffilmark{2}}
%\email{\emaila}

\altaffiltext{1}{Astronomy and Astrophysics Division, Physical Research Laboratory,
Navrangpura, Ahmedabad - 380 009, India. Email:asingal@prl.res.in}
\altaffiltext{2}{Thuringer Landessternwarte (TLS), Sternwarte 5, 07778 Tautenburg, Germany.
Email : kamlesh@tls-tautenburg.de}
\abstract{
We investigate the dependence of the Fanaroff--Riley (FR) 1/2 dichotomy of radio galaxies 
on their luminosities and redshifts. Because of a very strong 
redshift-luminosity correlation (Malmquist bias) in a flux-limited sample, any redshift-dependent effect 
could appear as a luminosity related effect and vice versa. 
A question could then arise -- do all the morphological differences seen in the two classes (FR 1 and 2 types) of sources,  
usually attributed to the differences in their luminosities, could as well be primarily a redshift-dependent 
effect? 
A sharp break in luminosity, seen among the two classes, could after all reflect a sharp redshift-dependence due to a 
rather critical ambient density value at some cosmic epoch. A doubt on these lines does not seem to have been raised 
in past and things have never been examined with this particular aspect in mind. 
We want to ascertain the customary prevalent view in the literature that the systematic differences in the 
two broad morphology types of FR 1 and 2 radio galaxies are indeed due to the differences in their luminosities, 
and not due to a change in redshift. 
Here we investigate the dependence of FR 1/2 dichotomy of radio galaxies on luminosity and redshift by using the 
3CR sample, where the FR 1/2 dichotomy was first seen, supplemented by data from an additional sample 
(MRC), that goes about a factor of 5 or more deeper in flux-density than the original 3CR sample. This 
lets us compare sources with similar luminosities but at different redshifts as well as examine sources at similar redshifts 
but with different luminosities, thereby allowing us a successful separation of the otherwise two intricately 
entangled effects. We find that the morphology type is not directly related to redshift and the break between the two types of 
morphologies seems to depend only upon the radio luminosity.
}
\maketitle
%\begin{flushright}
%\end{flushright}
\maketitle
\section{\bf INTRODUCTION}
One of the robust correlations in observational astronomy is between the morphology type of 
radio galaxies and their radio luminosity. First pointed out by Fanaroff and Riley (1974) 
that there is a very sharp dependence of the morphology type of radio galaxies on 
the luminosity so that almost all radio galaxies below a luminosity 
$P_{178} = 2 \times 10^{25}$ W Hz$^{-1}$ sr$^{-1}$ (for Hubble constant 
$H_{0}=50\,$km~s$^{-1}$\,Mpc$^{-1}$), are edge-darkened (called type 1) in their brightness distribution, 
while all radio galaxies above this luminosity limit are more or less edge-brightened (called type 2). 
This correlation has withstood the test of time (Miley 1980; Antonucci 1993, 2012; Urry and Padovani 1995; 
Kembhavi and Narlikar 1999). However, because of a very strong redshift-luminosity correlation (Malmquist bias) 
in a flux-limited sample, like in the 3CR sample used by Fanaroff and Riley (1974), 
any effect related with redshift could appear as a luminosity-dependent effect and vice versa. 
This then begs a question -- could all the morphological differences seen in FR 1 and 2 types of sources be primarily due to 
a transition across some critical redshift value, manifesting a cosmological evolutionary effect due to a critical ambient 
density value at that redshift, instead of, as almost universally believed, an effect of transition across a certain 
critical luminosity value? 

Following the archetypal paper by Fanaroff and Riley (1974), where they first pointed out the presence of 
two distinct morphology types of radio galaxies in the strong source 3CR sample, ascribing the distinct 
morphology of each galaxy to its radio luminosity, it has ever since been thought to be only a luminosity-dependent effect 
(see e.g., Saripalli 2012 and the references therein). 
There have been no attempts to investigate the alternative possibility that it could as well be a redshift-dependent 
effect, and thereby demonstrating the strong evolution of source morphology with cosmic epoch. 
For instance, suppose one wants to explore a correlation of the morphology type of radio galaxies with redshift. 
Figure 1 shows a scatter plot with redshift and luminosity of both types of morphologies classified by 
Fanaroff and Riley (1974) in their sample. 
The demarcation is as good with redshift as it is with luminosity. It might be a moot point to 
guess what would have been the verdict, had Fanaroff and Riley in their seminal paper  
tried a correlation against only redshift, instead of luminosity. It is quite likely the effect then would 
have been interpreted as due to a very strong cosmological evolution with redshift, and the subsequent theoretical interpretations 
in that case perhaps very different.

The two scenarios have very different physical interpretations, and either could be of equal importance.
For instance in the conventional interpretation, with intrinsic luminosities being the root cause of their different 
morphology types, there is a huge amount of literature about the relation between the luminosity break and 
the different  morphology types (Saripalli 2012 and the references therein). In fact all the models and discussions 
in the literature currently available are almost exclusively only within that framework. On the 
other hand a definite correlation with redshift alone would imply that it is the cosmic evolution of the properties 
of sources (perhaps because of the ambient density falling below a certain critical value due to the Hubble expansion) 
that might give rise to these two different type of morphologies, with FR 1 type being the current favourite   
and the FR 2 type ``a thing of past''. 
\begin{figure}[ht]
\includegraphics[width=\columnwidth]{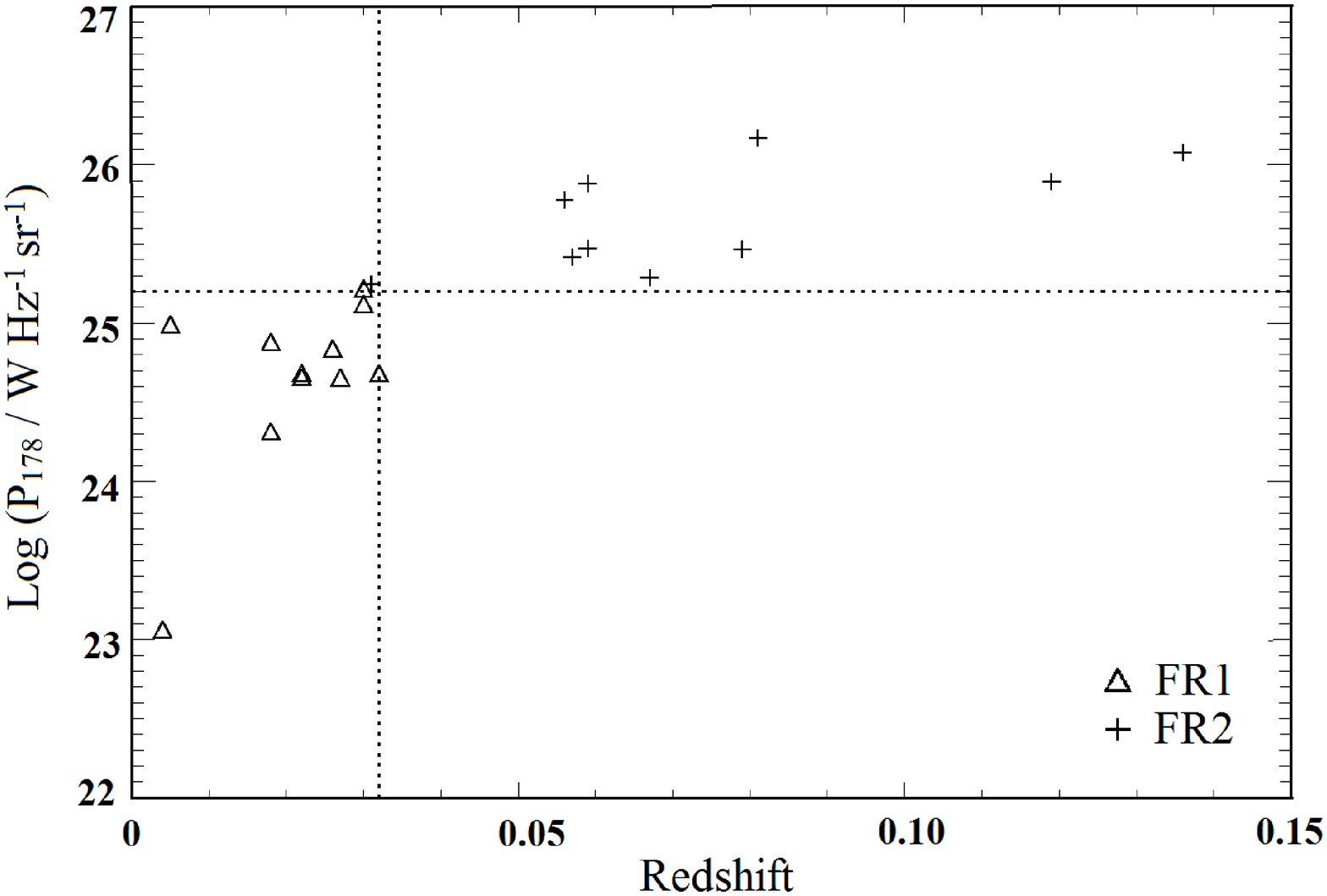}
\caption{A scatter plot of both types of FR morphologies (`$\Delta$' for edge-darkened FR 1 type and `$+$' for edge brightened 
FR 2 type) for the 3CR source sample of Fanaroff and Riley (1974) in the redshift-luminosity plane. To avoid undue 
compression of the scale, we have restricted the plot to a redshift limit of 0.15. There are no  
FR 1 type radio galaxies  which lie beyond a redshift of 0.032 or have a luminosity 
$P_{178} > 2 \times 10^{25}$ W Hz$^{-1}$ sr$^{-1}$ (for $H_{0}=50\,$km~s$^{-1}$\,Mpc$^{-1}$) 
in the sample of Fanaroff and Riley}
\end{figure}

All differences seen in FR 1 and FR 2 type sources, which are usually attributed to the differences  
in their luminosities, could as well be then related to the changing ambient densities with cosmic epoch. 
Also the very sharp division in luminosity could possibly be due to a critical ambient density 
value, which might divide the sources into two distinct morphology types. 
It might be the luminosity dependence or it might be the dependence on redshift that gives rise to these morphological 
differences, but this question could not be decisively settled based on any amount of arguments, sans actual observational 
data. At least this particular aspect has not yet been investigated in the literature. We may add that there are reports of 
FR 1 types seen at redshifts larger than $z>0.5$ (Saripalli et al. 2012),  
but a systematic investigation of this question is still needed using samples which 
are complete in the sense that all FR 1's above the sensitivity limit of the sample are included.
\section{The samples and the data}
To make these investigations one needs sufficient data comprising sources at different 
flux-density levels, so that one could examine sources with similar 
luminosities at different redshifts as well as compare them at similar redshifts for different luminosities, 
thereby separating the two effects. 
\begin{figure}[ht]
\includegraphics[width=6cm]{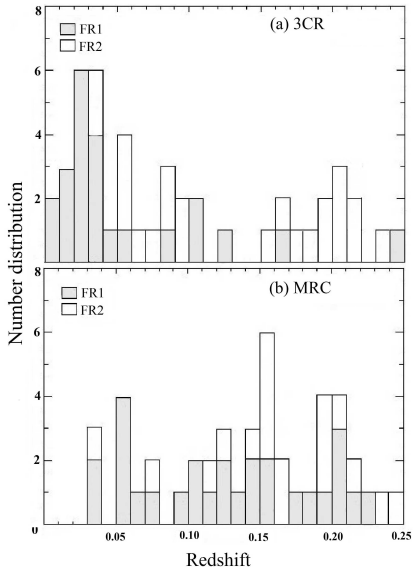}
\caption{Histograms showing distributions of FR 1/2 morphology types radio galaxies with redshift
for the (a) 3CR and (b) MRC samples. 
It should be noted that all radio galaxies seen beyond the redshift limit of the plots 
(i.e. $z > 0.25$) are only of FR 2 type in both samples}
\end{figure}
\begin{figure}[ht]
\includegraphics[width=6.1cm]{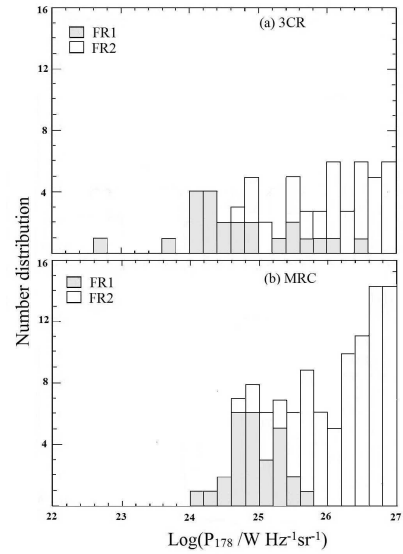}
\caption{Histograms showing distributions of FR 1/2 morphology types radio galaxies  with radio luminosity 
($P_{178}$ in units of W Hz$^{-1}$ sr$^{-1}$) for the (a) 3CR and (b) MRC samples. 
It should be noted that all radio galaxies seen beyond the luminosity limit of the plots 
(i.e. $P_{178} > 10^{27}$ W Hz$^{-1}$ sr$^{-1}$) are only of FR 2 type 
in both samples}
\end{figure}
We investigate this dependence of  FR dichotomy of radio galaxies on luminosity and redshift 
by taking data from two such different samples. 
And since the transition value of luminosity or/and redshift 
may not be as sharp (see e.g., Baum et al. 1995 and the references 
therein) as inferred from the data used by Fanaroff and Riley (1974), we investigate the two effects by 
determining the median values of luminosity and redshift for FR 1 type sources in each one of our samples. 

Our first sample is the 3CR (Laing et al. 1983), which is a complete strong source sample, 
with all necessary optical and radio information with good resolution maps so that one can in most cases 
unambiguously decide the FR 1/2 type of morphologies. The sample is selected to include all radio sources with low-frequency 
(178 MHz) flux-density $S_{178} > 10.9$ Jy with declination $\delta > 10^\circ$ and the galactic latitude $b > 10^\circ$. 
The sample covers a solid angle of 4.23 steradian and contains a total of 173 sources. The updated data for the Laing et al. (1983) 
sample are available at {\em http://astroherzberg.org/people/chris-willott/research/3crr/}.

The second sample we have chosen is the essentially complete MRC (Molonglo Reference Catalog) 
sample (Kapahi et al. 1998) with $S_{408} \ge 0.95$ Jy in the declination range $-30^\circ < \delta < -20^\circ, b > 20^\circ$ 
but excluding the RA range $14^h03^m - 20^h 20^m$. The MRC sample is about a factor $\sim 5$ or more deeper than the 3CR sample and 
has the required radio and optical information. 
The total sample comprises 550 sources, with 111 of them being quasars and the 
remainder radio galaxies. Optical identifications for the latter are complete up to a red magnitude of $\sim 24$ 
or a {\em K} magnitude of  $\sim 19$ and among the still unidentified ones, which are expected to be at high redshifts 
$z \stackrel{>}{_{\sim}}1$ and therefore of high luminosities as well, it is unlikely that there would be many FR 1 types.

\begin{table*}[ht]
\caption{Percentile values of redshift and luminosity distributions for the FR 1 sources in the two samples.}
%\begin{tabular}{@{}lccccccccc}
\begin{tabular}{@{\hspace{0mm}}l@{\hspace{3mm}}c@{\hspace{3mm}}c@{\hspace{3mm}}c@{\hspace{3mm}}c
@{\hspace{3mm}}c@{\hspace{4mm}}c@{\hspace{4mm}}c@{\hspace{4mm}}c}
\hline
Sample & $S_{\rm 178,med}$ & Number & $z_{\rm med}$ & $z_{\rm lq}$ &$z_{\rm uq}$ & 
log($P_{\rm 178,med}$) & log($P_{\rm 178,lq}$) & log($P_{\rm 178,uq}$) \\
& (Jy) & FR 1 & & && (W Hz$^{-1}$sr$^{-1}$) &(W Hz$^{-1}$sr$^{-1}$) &(W Hz$^{-1}$sr$^{-1}$) \\
%& & & & & &  \\
\hline
3CR & $19.4$ & 23 & $0.03\pm 0.005$ & 0.02 & 0.08 & $24.6\pm 0.2$  & 24.2 &    25.5      \\
MRC & $2.7$ & 27 & $0.12\pm 0.015$ & 0.06 & 0.17 & $24.9\pm 0.1$  & 24.6 &    25.2      \\
\hline
\end{tabular}
\end{table*}

To quantitatively distinguish between  FR 1 and 2, following Fanaroff and Riley (1974), we classify a radio galaxy 
as FR 1 if the separation between the points of peak intensity 
in the two lobes is smaller than half the largest size of the source. Similarly FR 2 is the one in which the 
separation between the points of peak 
intensity in the two lobes is greater than half the largest size of the source. 
This is equivalent to having the ``hot spots'' nearer to (FR 1) or further away from (FR 2) 
the central optical galaxy than the regions of diffuse radio emission.
In both samples we have examined the radio maps and using the above criteria, we have classified each source  
into either of the two types (FR 1 and 2). There are a small number of sources that have an ambiguous FR classification. 
For example, there are sources of a hybrid type called HYMORS (one lobe of FR1 and the second of FR2 type (see, e.g., 
Gopal-Krishna and Wiita 2000).  Then there are some radio galaxies exhibiting two, or even three, pairs of lobes, 
where the AGN jet activity may have occurred more than once during a lifetime of a parent galaxy. These
radio sources are called double-double radio galaxies (DDRGs), with some clear examples given by 
Saripalli, Subrahmanyan and Udaya Shankar (2002, 2003; also see Saikia and Jamrozy 2009). We excluded such cases 
where there was ambiguity in classifying them as FR1 or FR2. The number of such sources being relatively 
very small, it should not be too detrimental to our conclusions.

The luminosity of a source in our sample is calculated from its flux density $S_{178}$ and the spectral index 
$\alpha $ ($S\propto \nu^{-\alpha}$) as 
\begin{eqnarray}
P_{178}=S_{178}{\cal D}^2(1+z)^{1+\alpha},
\end{eqnarray}
where ${\cal D}$ is the comoving cosmological distance calculated from the cosmological redshift $z$ of the source. 
In general it is not possible to express ${\cal D}$ in terms of 
$z$ in a close-form analytical expression and one may have to evaluate it 
numerically. For example, in the  flat universe models   
($\Omega_m+\Omega_\Lambda=1, \Omega_\Lambda \neq 0$), ${\cal D}$ is given by (see e.g., Weinberg 2008),
\begin{eqnarray}
{\cal D}=\frac{c}{H_0}\int^{1+z}_{1}\frac{{\rm d}z}{\left(\Omega_\Lambda+\Omega_m z^3\right)^{1/2}}\;.
\end{eqnarray}
For a given $\Omega_\Lambda$, ${\cal D}$ can be evaluated from Eq. (2) by a numerical integration. 
Here $\Omega_m$ is the matter energy density (including that of the dark matter) and $\Omega_\Lambda$ is the vacuum energy 
(dark energy!) density, both defined in terms of the critical energy density $\Omega_{\rm c}= 3 H_{0}^2c^2/(8\pi G)$, where $G$ is 
the gravitational constant and $c$ is the velocity of light in vacuum.   
We have used $H_{0}=71\,$km~s$^{-1}$\,Mpc$^{-1}$, $\Omega_m=0.27$ and $\Omega_\Lambda=0.73$ (Spergel et al. 2003). 

The detailed data used by us are given in Appendix, listed separately for both samples in Tables 2 and 3, which are self-explanatory.
\section{Results and discussion}
Figure 2 shows histograms of the distributions of radio galaxies with FR 1 type morphology against redshift for both our 
samples. We have restricted our plots to $z=0.25$ only, as    
no FR 1 type radio galaxy is seen beyond this redshift in either sample. 
We have determined median value $z_{med}$ for the redshift distribution of FR 1 sources for each sample. The
details of the method for determining the median values and for the
estimation of their rms errors are described in Singal~(1988).

Figure 3 shows similar histograms of the distribution of radio galaxies with FR 1 type morphology against 
luminosity (at 178 MHz) for our two samples. 
Since all the FR 1's we are interested in are at low redshifts 
($z<0.25$), the specific cosmological parameters do not make much difference for the luminosity evaluations except for 
the scaling factor due to the Hubble constant, which in our case makes luminosity estimates lower by a factor of 
$(71/50)^2 \approx 2$ than the calculations in Fanaroff and Riley (1974). Thus the FR 1/2 luminosity break of 
$P_{178} = 2 \times 10^{25}$ W Hz$^{-1}$ sr$^{-1}$ (for $H_{0}=50\,$km~s$^{-1}$\,Mpc$^{-1}$), 
arrived at by Fanaroff and Riley (1974) in their seminal paper, corresponds to 
$P_{178} = 10^{25}$ W Hz$^{-1}$ sr$^{-1}$  in our case. 
Again we have restricted our plots to $P_{178}=10^{27}$ W Hz$^{-1}$ sr$^{-1}$ only, as    
no FR 1 type radio galaxy is seen beyond this luminosity value in either sample. 

From Figure 3, it is also clear that the FR 1/2 break is not as sharp as stated in Fanaroff and Riley (1974), 
as some overlap of both type of morphologies is seen in luminosity. However there is no denying that in all samples 
there are only a few, if any, FR 1 types with luminosities $P_{178}>10^{26}$ W Hz$^{-1}$ sr$^{-1}$. 

In Table 1 we have 
listed the median values both for the redshift and the luminosity in each sample. 
Also listed are the median flux-density of FR 1's in each sample and the number of FR 1 sources in the sample. 
The rms error in $z_{med}$ in each case is determined from the frequency distribution (histogram) of 
$z$-distribution. 
The rms error is given by $\sqrt {n}/(2 f_p)$ in units of the class interval of $z$ (Kendall 1945; Yule and Kendall 1950), 
where $n$ is the total number of sources in the sample and $f_p$ is the ordinate value of the smoothed frequency 
distribution at the median value in Figure 2. 
The median value of the distribution does seem to shift substantially with redshift for samples which differ 
in the median flux-density by a factor of $\sim 7$. $P_{med}$ and the rms error in $P_{med}$ in each case is determined from 
the frequency distribution (histogram) of $P$-distribution in Figure 3, in the same way as for $z_{med}$ as described 
above. 
Within errors there is hardly any difference in the two samples in the $P_{med}$ value, which thus seems to be independent 
of the flux-density level of the sample. 
\begin{figure}[ht]
\includegraphics[width=\columnwidth]{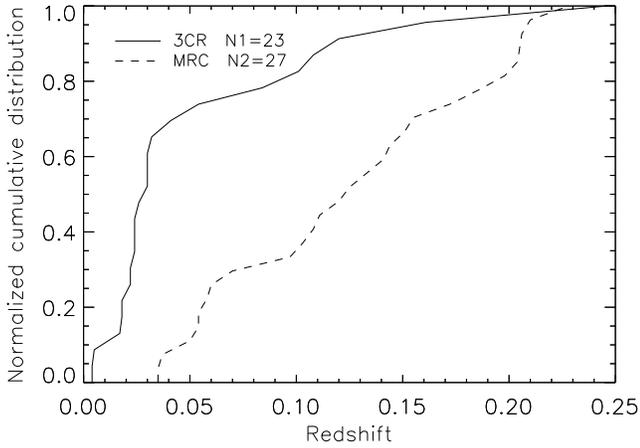}
\caption{Normalized cumulative distributions of FR 1 morphology type radio galaxies  with redshift for the 
(a) 3CR sample (continuous curve) and (b) MRC sample (broken curve). N1 and N2 give the number of FR 1 radio galaxies in the 
3CR and MRC samples, respectively}
\end{figure}
While the median value of redshift ($z_{\rm med}$) for the weaker sample as compared to that of the stronger 3CR sample, differs by 
as much as a factor of about four, at about a $5\sigma$ level, the difference in the median value of luminosity ($P_{\rm med}$) 
is only marginal, being only a factor of about $10^{0.3} \sim 2$ and that too within only about $1\sigma$.
From $z_{\rm med}$ and log($P_{\rm med}$) values, it is clear 
that the FR 1 type of morphology of  radio galaxies is indeed due to their luminosity below a critical value 
as indeed envisaged first time by Fanaroff and Riley (1974) and that it is not directly related to the redshift and hence 
not due to a cosmic evolution effect. 

To ascertain it further we have examined the normalized cumulative distributions of FR 1 morphology type radio galaxies with redshift 
as well as luminosity. Figure 4 shows the normalized cumulative distributions of FR 1 morphology type radio galaxies with redshift 
for the two samples. Since the cumulative distributions shown are the normalized ones, therefore it automatically accounts for the fact  
that the sky coverages in the 3CR and MRC samples are different and we can directly compare the relative 
distributions of FR 1 types up to any given redshift, and thereby for the two corresponding different radio 
luminosity values in the two samples. If the morphology type depended only on redshift and 
not on luminosity, then both normalized cumulative distributions in Figure 4 should be more or less coincident, 
which definitely is not the case. In fact a Kolmogorov-Smirnov (K-S) test shows that the null-hypothesis that 
the two distribution are the same can be ruled out at a $99.99\%$ confidence level.
\begin{figure}[ht]
\includegraphics[width=\columnwidth]{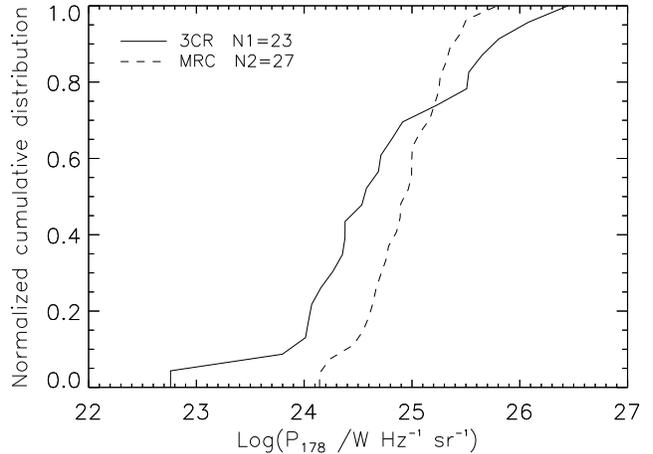}
\caption{Normalized cumulative distributions of FR 1 morphology type radio galaxies  with radio luminosity 
($P_{178}$ in units of W Hz$^{-1}$ sr$^{-1}$) for the 
(a) 3CR sample (continuous curve) and (b) MRC sample (broken curve). N1 and N2 give the number of FR 1 radio galaxies in the 
3CR and MRC samples, respectively}
\end{figure}

In Table 1 we have listed, along with the median value ($z_{\rm med}$), also the
lower quartile ($z_{\rm lq}$) and upper quartile ($z_{\rm uq}$) of the cumulative distribution of the redshifts   
of the sources in these samples. We see that the space distribution of FR 1 morphology type radio galaxies 
does change with the flux-density levels of the sample. 
There is a large difference  
in the redshift distribution of the 3CR sample from those of the MCR sample. Note that the 
3CR sample is about 7 times stronger in flux-density than the MRC sample. It seems that the redshift 
distribution of the FR 1 radio galaxies does depend on the flux-density level of the sample. Of course one may have to 
examine still weaker samples, where one might see FR 1 types $z>0.25$, to see the generality of these results.

Figure 5 shows the normalized cumulative distribution of radio luminosity for FR 1 morphology type radio galaxies 
for both samples. The distributions overlap and there seems to be no gross difference in the luminosity distribution   
of FR 1 sources in the two samples. In Table 1 we have also listed the 
lower quartile ($P_{\rm lq}$) and upper quartile ($P_{\rm uq}$) of the cumulative distribution of the radio luminosity 
of the source in the two samples, which again are of very similar values. From these  
we find that half of FR 1 types lie in the narrow range of luminosities 
$P_{178} \sim 10^{24.7\pm 0.5}$ W Hz$^{-1}$ sr$^{-1}$, that is within a factor of 3 around $5 \times 10^{24}$ 
W Hz$^{-1}$ sr$^{-1}$. Also all galaxies with radio luminosity above $P_{178} \sim 10^{26.5}$ W Hz$^{-1}$ sr$^{-1}$ 
are only of type 2 with an edge brightened morphology. 
We have thus reaffirmed that the FR 1/2 dichotomy is due to a change in luminosity below and above a 
certain critical level, as first proposed by Fanaroff and Riley (1974) and followed in the literature 
ever since, and that it is not due to any cosmic epoch dependent evolution that gives rise to different FR 1 and 2 type morphologies. 

Here we may add that the dual-population unification scheme (see e.g., Jackson \& Wall 1999), where FR 1 and 2 type radio galaxies 
form two intrinsically distinct `unbeamed' or `side-on' parent populations of the steep spectrum extragalactic radio sources, has been 
running into serious 
problems as far as the FR 2 type radio galaxy part is concerned. Unacceptably large mismatches with the predictions of the relative 
number and radio-size distributions of FR 2 radio galaxies and extended steep-spectrum quasars are seen in various redshift bins 
in different unbiased radio complete samples that have been examined (Singal \& Singh 2013a, 2013b). There is almost complete absence of predicted 
foreshortening in the quasar radio sizes due to projection effects over and above the statistical spread. In the 3CR sample too, 
the predictions of unification scheme are not corroborated by the observed radio-size data even when the low-excitation radio 
galaxies from the FR 2 sample are excluded (Singal 2014). Except for that particular bin ($0.5 \le z <1$) of the 3CR sample, which 
incidentally was instrumental in the original proposition of the unification scheme (Barthel 1989), data in other 
redshift bins do not seem to yield the expected size ratios of FR 2 radio galaxies and quasars. 
It of course remains to be seen if the FR 1 type radio galaxy part of the dual-population unification scheme still holds true.

\section{Conclusions}
We have investigated if there is a direct dependence of the FR 1 and 2 morphology types of the radio galaxies on redshift. 
For this we compared their distributions in two different samples with different flux-density limits, which allowed 
us to separately study the effects of redshift and/or radio luminosity on the occurrences of the two morphology types.  
It was shown that the morphology type is not directly related to redshift and thereby not a cosmic epoch dependent effect. 
The break between the two types of morphologies seems to depend only upon the radio luminosity.  Half of the 
FR 1 type radio galaxies lie in the narrow range of luminosities 
$P_{178} \sim 10^{24.7\pm 0.5}$ W Hz$^{-1}$ sr$^{-1}$, with none exceeding the value
$P_{178} \sim 10^{26.5}$ W Hz$^{-1}$ sr$^{-1}$, above which all were found to be exclusively type 2 with 
an edge brightened morphology. 
\section*{ACKNOWLEDGEMENTS}
KR expresses her gratitude to the Astronomy and Astrophysics Division  
of the Physical Research laboratory Ahmedabad, where work on this summer project was done under the 
guidance of AKS. 
\section*{Appendix}
Tables 2 and 3 give the radio and optical data used by us for the 3CR and MRC samples.
\clearpage
\onecolumn
\begin{longtable}{@{}lllclc}
\caption{Radio and optical data for the 3CR sample}\\
%\tiny
\hline
\,\,Source& $S_{178}$ & \,\,$\alpha$ & \,\,\,FR & \,\,\,\,$z$ & $\log(P_{178})$ \\
\,\,Name& (Jy)& \,\,\,\,\,\,&\,\,\,Type & & (W Hz$^{-1}$sr$^{-1}$)\\
\,\,\,\,(1)&(2)&\,(3)&(4)&\,\,(5)&(6)\\
\hline
\endfirsthead
\tablename ~2\ -- \textit{continued} \\\\
\hline
\,\,Source& $S_{178}$ & \,\,$\alpha$ & \,\,\,FR & \,\,\,\,$z$ & $\log(P_{178})$ \\
\,\,Name& (Jy)& \,\,\,\,\,\,&\,\,\,Type & & (W Hz$^{-1}$sr$^{-1}$)\\
\,\,\,\,(1)&(2)&\,(3)&(4)&\,\,(5)&(6)\\
\hline
\endhead
4C12.03 	&	10.9	&	0.87	&	FR2	&	0.156	&	25.74	\\
3C16 	&	12.2	&	0.954	&	FR2	&	0.406	&	26.74	\\
3C19 	&	13.18	&	0.637	&	FR2	&	0.482	&	26.90	\\
3C20 	&	46.76	&	0.606	&	FR2	&	0.174	&	26.46	\\
3C28 	&	17.76	&	1.011	&	FR2	&	0.195	&	26.18	\\\\
3C31 	&	18.31	&	0.682	&	FR1	&	0.018	&	24.01	\\
3C33 	&	59.29	&	0.701	&	FR2	&	0.059	&	25.58	\\
3C35 	&	11.44	&	0.907	&	FR2	&	0.067	&	24.98	\\
3C42 	&	13.08	&	0.705	&	FR2	&	0.395	&	26.71	\\
3C46 	&	11.11	&	0.905	&	FR2	&	0.437	&	26.77	\\\\
3C61.1 	&	34	&	0.736	&	FR2	&	0.188	&	26.40	\\
3C66B 	&	26.81	&	0.736	&	FR1	&	0.022	&	24.35	\\
3C76.1 	&	13.29	&	0.588	&	FR1	&	0.032	&	24.38	\\
3C83.1B 	&	28.99	&	0.649	&	FR1	&	0.026	&	24.54	\\
3C84 	&	66.81	&	1.141	&	FR1	&	0.018	&	24.58	\\\\
3C98 	&	51.44	&	0.732	&	FR2	&	0.031	&	24.94	\\
4C14.11 	&	12.09	&	0.84	&	FR2	&	0.207	&	26.05	\\
3C132 	&	14.93	&	0.79	&	FR2	&	0.214	&	26.17	\\
3C153 	&	16.67	&	0.577	&	FR2	&	0.277	&	26.45	\\
3C173.1 	&	16.78	&	0.898	&	FR2	&	0.292	&	26.54	\\\\
DA240 	&	23.21	&	0.77	&	FR2	&	0.035	&	24.70	\\
3C192 	&	22.99	&	0.81	&	FR2	&	0.059	&	25.17	\\
3C200 	&	12.31	&	0.829	&	FR2	&	0.458	&	26.85	\\
4C14.27 	&	11.22	&	1.15	&	FR2	&	0.392	&	26.70	\\
4C73.08 	&	15.58	&	0.85	&	FR2	&	0.0581	&	24.98	\\\\
3C264 	&	28.34	&	0.82	&	FR1	&	0.022	&	24.38	\\
3C272.1 	&	21.14	&	0.6	&	FR1	&	0.004	&	22.76	\\
A1552 	&	12.53	&	0.94	&	FR1	&	0.084	&	25.23	\\
3C274 	&	1144.5	&	0.792	&	FR1	&	0.005	&	24.69	\\
3C274.1 	&	17.98	&	0.936	&	FR2	&	0.422	&	26.95	\\\\
3C284 	&	12.31	&	0.889	&	FR2	&	0.239	&	26.21	\\
3C285 	&	12.31	&	0.786	&	FR2	&	0.079	&	25.16	\\
3C288 	&	20.6	&	0.775	&	FR1	&	0.246	&	26.45	\\
3C296 	&	14.17	&	0.745	&	FR1	&	0.024	&	24.16	\\
3C299 	&	12.86	&	0.557	&	FR2	&	0.367	&	26.61	\\\\
3C300 	&	19.51	&	0.837	&	FR2	&	0.27	&	26.52	\\
3C305 	&	17.11	&	0.816	&	FR1	&	0.041	&	24.71	\\
3C310 	&	60.05	&	0.974	&	FR1	&	0.054	&	25.51	\\
3C314.1 	&	11.55	&	1.023	&	FR1	&	0.12	&	25.53	\\
3C315 	&	19.4	&	0.885	&	FR1	&	0.108	&	25.65	\\\\
3C319 	&	16.67	&	0.852	&	FR2	&	0.192	&	26.12	\\
3C321 	&	14.71	&	0.825	&	FR2	&	0.096	&	25.42	\\
3C326 	&	22.23	&	0.88	&	FR2	&	0.088	&	25.52	\\
NGC6109 	&	11.66	&	0.76	&	FR1	&	0.03	&	24.27	\\
3C338 	&	51.12	&	1.074	&	FR1	&	0.03	&	24.91	\\\\
3C341 	&	11.77	&	0.863	&	FR2	&	0.448	&	26.81	\\
NGC6251 	&	10.9	&	0.72	&	FR1	&	0.024	&	24.04	\\
3C346 	&	11.88	&	0.807	&	FR1	&	0.161	&	25.80	\\
3C349 	&	14.49	&	0.739	&	FR2	&	0.205	&	26.12	\\
3C381 	&	18.09	&	0.729	&	FR2	&	0.16	&	25.98	\\\\
3C388 	&	26.81	&	0.683	&	FR2	&	0.09	&	25.61	\\
3C401 	&	22.78	&	0.635	&	FR2	&	0.201	&	26.29	\\
3C433 	&	61.25	&	0.719	&	FR1	&	0.101	&	26.08	\\
3C436 	&	19.4	&	0.855	&	FR2	&	0.214	&	26.29	\\
3C438 	&	48.72	&	0.822	&	FR2	&	0.29	&	26.99	\\\\
3C449 	&	12.53	&	0.742	&	FR1	&	0.017	&	23.80	\\
3C452 	&	59.29	&	0.825	&	FR2	&	0.081	&	25.87	\\
NGC7385 	&	11.66	&	0.75	&	FR1	&	0.024	&	24.07	\\
3C457 	&	14.27	&	1.229	&	FR2	&	0.428	&	26.91	\\
3C465 	&	41.2	&	0.833	&	FR1	&	0.03	&	24.82	\\
\hline
\end{longtable}

%\pagebreak
\clearpage
\begin{longtable}{@{}lllclc}
%\hline\hline
%\begin{longtable}{@{\hspace{0mm}}c@{\hspace{3mm}}c@{\hspace{3mm}}c@{\hspace{3mm}}c@{\hspace{3mm}}
%c@{\hspace{3mm}}c@{\hspace{3mm}}c@{\hspace{3mm}}c@{\hspace{3mm}}c}}}
\caption{Radio and optical data for the MRC sample}\\
%\tiny
\hline
\,\,Source& $S_{408}$ & \,\,$\alpha$ & \,\,\,FR & \,\,\,\,$z$ & $\log(P_{178})$ \\
\,\,Name& (Jy)& \,\,\,\,\,\,&\,\,\,Type & & (W Hz$^{-1}$sr$^{-1}$)\\
\,\,\,\,(1)&(2)&\,(3)&(4)&\,\,(5)&(6)\\
\hline
\endfirsthead
\tablename ~3\ -- \textit{continued} \\\\
\hline
\,\,Source& $S_{408}$ & \,\,$\alpha$ & \,\,\,FR & \,\,\,\,$z$ & $\log(P_{178})$ \\
\,\,Name& (Jy)& \,\,\,\,\,\,&\,\,\,Type & & (W Hz$^{-1}$sr$^{-1}$)\\
\,\,\,\,(1)&(2)&\,(3)&(4)&\,\,(5)&(6)\\
\hline
\endhead
B0001-233	&	1.23	&	0.99	&	FR1	&	0.097	&	24.71	\\
B0001-237	&	1.77	&	0.83	&	FR2	&	0.315	&	25.93	\\
B0017-205	&	1.96	&	0.78	&	FR2	&	0.197	&	25.49	\\
B0020-253	&	5.36	&	0.78	&	FR2	&	0.35	&	26.49	\\
B0022-209	&	1.19	&	0.94	&	FR1	&	0.054	&	24.14	\\\\
B0028-223	&	1.18	&	0.79	&	FR1	&	0.205	&	25.32	\\
B0048-233	&	1.18	&	0.84	&	FR1	&	0.111	&	24.76	\\
B0055-256	&	1.18	&	0.97	&	FR2	&	0.199	&	25.37	\\
B0115-261	&	2.58	&	0.74	&	FR2	&	0.268	&	25.89	\\
B0125-216	&	1.29	&	0.81	&	FR2	&	0.34	&	25.86	\\\\
B0137-263	&	1.46	&	0.83	&	FR2	&	0.16	&	25.19	\\
B0143-246	&	1.51	&	0.86	&	FR2	&	0.716	&	26.72	\\
B0146-224	&	1.65	&	0.81	&	FR2	&	0.36	&	26.02	\\
B0148-297	&	7.04	&	0.82	&	FR2	&	0.41	&	26.79	\\
B0149-260	&	1.05	&	0.95	&	FR1	&	0.144	&	25.00	\\\\
B0149-299	&	2.42	&	0.83	&	FR2	&	0.603	&	26.72	\\
B0155-212	&	2.39	&	1.04	&	FR2	&	0.159	&	25.49	\\
B0205-229	&	1.77	&	0.83	&	FR2	&	0.68	&	26.71	\\
B0208-240	&	1.87	&	0.74	&	FR2	&	0.23	&	25.60	\\
B0209-282	&	1.56	&	0.81	&	FR2	&	0.6	&	26.52	\\\\
B0226-284	&	1.063	&	0.69	&	FR1	&	0.21	&	25.25	\\
B0245-297	&	1.16	&	0.75	&	FR2	&	0.36	&	25.84	\\
B0305-226	&	4.95	&	0.88	&	FR2	&	0.268	&	26.24	\\
B0313-271	&	1.8	&	1.14	&	FR2	&	0.216	&	25.70	\\
B0325-260	&	1.04	&	0.74	&	FR2	&	0.638	&	26.37	\\\\
B0326-288	&	4.03	&	0.83	&	FR1	&	0.108	&	25.26	\\
B0346-297	&	1.72	&	0.88	&	FR2	&	0.413	&	26.21	\\
B0357-247	&	2.16	&	0.87	&	FR2	&	0.205	&	25.61	\\
B0412-204	&	2.51	&	1	&	FR2	&	0.69	&	26.98	\\
B0424-268	&	3.25	&	0.87	&	FR2	&	0.47	&	26.62	\\\\
B0428-271	&	1.72	&	0.83	&	FR2	&	0.84	&	26.92	\\
B0428-281	&	2.49	&	0.89	&	FR2	&	0.65	&	26.85	\\
B0430-235	&	0.96	&	0.86	&	FR2	&	0.82	&	26.66	\\
B0436-294	&	1.2	&	0.99	&	FR2	&	0.808	&	26.82	\\
B0442-282	&	18.85	&	0.97	&	FR2	&	0.147	&	26.28	\\\\
B0452-260	&	0.98	&	0.83	&	FR1	&	0.141	&	24.90	\\
B0453-206	&	11.25	&	0.7	&	FR1	&	0.035	&	24.64	\\
B0457-247	&	1.25	&	0.7	&	FR1	&	0.186	&	25.21	\\
B0503-284	&	9	&	0.9	&	FR1	&	0.037	&	24.67	\\
B0508-220	&	5.1	&	0.81	&	FR2	&	0.16	&	25.72	\\\\
B0512-200	&	0.98	&	0.69	&	FR1	&	0.133	&	24.78	\\
B0522-239	&	1.08	&	0.8	&	FR2	&	0.5	&	26.17	\\
B0529-210	&	1.33	&	0.96	&	FR2	&	0.42	&	26.16	\\
B0541-243	&	3.61	&	0.98	&	FR2	&	0.523	&	26.83	\\
B0602-289	&	1.37	&	1.03	&	FR2	&	0.56	&	26.51	\\\\
B0938-205	&	1.35	&	0.88	&	FR2	&	0.371	&	26.00	\\
B0952-224	&	1.71	&	0.62	&	FR1	&	0.228	&	25.50	\\
B0959-225	&	1.04	&	0.68	&	FR2	&	0.895	&	26.67	\\
B0959-263	&	1.72	&	0.82	&	FR2	&	0.677	&	26.69	\\
B1006-214	&	1.35	&	0.78	&	FR2	&	0.246	&	25.55	\\\\
B1006-286	&	3.54	&	0.76	&	FR2	&	0.582	&	26.81	\\
B1023-226	&	1.08	&	0.94	&	FR2	&	0.586	&	26.41	\\
B1025-270	&	1.82	&	0.83	&	FR2	&	0.72	&	26.79	\\
B1026-202	&	1.95	&	0.88	&	FR2	&	0.566	&	26.59	\\
B1027-225	&	1.11	&	1.1	&	FR2	&	0.15	&	25.12	\\\\
B1029-233	&	1.08	&	0.86	&	FR2	&	0.611	&	26.41	\\
B1033-251	&	1.53	&	0.83	&	FR2	&	0.44	&	26.20	\\
B1036-215	&	0.98	&	0.82	&	FR2	&	0.585	&	26.30	\\
B1048-238	&	1.31	&	0.76	&	FR1	&	0.206	&	25.35	\\
B1056-272	&	0.97	&	0.97	&	FR2	&	0.25	&	25.50	\\\\
B1103-244	&	3.9	&	0.97	&	FR1	&	0.05	&	24.60	\\
B1126-246	&	1.1	&	0.74	&	FR1	&	0.155	&	25.00	\\
B1126-290	&	2.42	&	0.74	&	FR2	&	0.41	&	26.28	\\
B1129-250	&	0.98	&	0.91	&	FR2	&	1.065	&	26.98	\\
B1136-211	&	1.1	&	1.02	&	FR2	&	0.87	&	26.88	\\\\
B1222-252	&	1.82	&	0.99	&	FR2	&	0.077	&	24.67	\\
B1226-211	&	3.28	&	0.8	&	FR2	&	0.191	&	25.70	\\
B1240-209	&	4.58	&	0.87	&	FR2	&	0.42	&	26.65	\\
B1251-289	&	4.22	&	1.2	&	FR1	&	0.058	&	24.86	\\
B1257-253	&	1.5	&	0.68	&	FR1	&	0.06	&	24.24	\\\\
B1303-215	&	1.46	&	0.76	&	FR1	&	0.12	&	24.89	\\
B1309-211	&	1.67	&	0.83	&	FR2	&	0.3	&	25.86	\\
B1313-248	&	2.14	&	0.9	&	FR2	&	0.74	&	26.93	\\
B1325-257	&	1.16	&	0.93	&	FR2	&	0.62	&	26.49	\\
B1343-253	&	1.05	&	0.74	&	FR2	&	0.129	&	24.81	\\\\
B1346-252	&	1.27	&	1.26	&	FR1	&	0.125	&	25.07	\\
B1358-214	&	1.48	&	1.08	&	FR2	&	0.5	&	26.45	\\
B2020-211	&	1.59	&	1.43	&	FR1	&	0.054	&	24.46	\\
B2038-280	&	1.47	&	1.23	&	FR2	&	0.39	&	26.26	\\
B2039-236	&	1.63	&	0.86	&	FR2	&	0.621	&	26.60	\\\\
B2040-219	&	1.2	&	1.08	&	FR1	&	0.204	&	25.45	\\
B2045-245	&	1.99	&	1	&	FR2	&	0.73	&	26.94	\\
B2053-201	&	6.37	&	0.8	&	FR2	&	0.155	&	25.79	\\
B2057-286	&	2.34	&	0.93	&	FR2	&	0.605	&	26.77	\\
B2101-214	&	0.96	&	0.73	&	FR1	&	0.198	&	25.17	\\\\
B2104-256	&	28.1	&	0.79	&	FR2	&	0.037	&	25.12	\\
B2116-250	&	1.74	&	0.89	&	FR2	&	0.467	&	26.35	\\
B2117-269	&	2.6	&	0.77	&	FR1	&	0.103	&	25.00	\\
B2118-266	&	1.17	&	0.58	&	FR2	&	0.343	&	25.71	\\
B2132-236	&	0.95	&	0.95	&	FR2	&	0.81	&	26.70	\\\\
B2134-281	&	2.12	&	0.73	&	FR1	&	0.07	&	24.55	\\
B2137-279	&	1.21	&	0.94	&	FR2	&	0.64	&	26.55	\\
B2148-228	&	1.42	&	0.99	&	FR2	&	0.85	&	26.95	\\
B2206-251	&	2.04	&	0.93	&	FR2	&	0.158	&	25.36	\\
B2226-224	&	1.25	&	0.89	&	FR2	&	0.38	&	26.00	\\\\
B2236-264	&	1.23	&	0.83	&	FR2	&	0.43	&	26.08	\\
B2254-248	&	1.81	&	0.68	&	FR2	&	0.54	&	26.40	\\
B2256-207	&	1.21	&	1.02	&	FR2	&	0.87	&	26.93	\\
B2308-214	&	0.99	&	0.83	&	FR1	&	0.151	&	24.97	\\
B2311-222	&	2.24	&	0.78	&	FR2	&	0.434	&	26.33	\\\\
B2313-277	&	1.9	&	0.97	&	FR2	&	0.614	&	26.72	\\
B2317-277	&	5.44	&	0.73	&	FR1	&	0.173	&	25.79	\\
B2321-228	&	1.03	&	1.21	&	FR2	&	0.114	&	24.87	\\
B2324-259	&	1.44	&	0.72	&	FR2	&	0.286	&	25.69	\\
B2325-213	&	3.07	&	0.94	&	FR2	&	0.58	&	26.85	\\\\
B2343-243	&	1.85	&	0.8	&	FR2	&	0.6	&	26.59	\\
B2348-235	&	1.47	&	0.83	&	FR2	&	0.952	&	26.98	\\
\hline
\end{longtable}
%\clearpage
\twocolumn

%\nopagebreak
\end{document}